\documentclass[preprintnumbers,amsmath,amssymb,superscriptaddress]{revtex4-2}

\usepackage{mathrsfs}
\usepackage{amssymb}
\usepackage{graphicx}
\usepackage{dcolumn}
\usepackage{bm}
\usepackage{textcomp}
\usepackage{amsmath}
\usepackage[titletoc]{appendix}
\usepackage{accents}
\usepackage{ntheorem}
\usepackage{color}

\newcommand\ket[1]{\ensuremath{|#1\rangle}}

\newcommand\iprod[2]{\ensuremath{\langle#1|#2\rangle}}

\newcommand\tr{\mathop{\rm tr}\nolimits}
\newcounter{RomanNumber}

\makeatletter
\def\widebar{\accentset{{\cc@style\underline{\mskip10mu}}}}
\def\Widebar{\accentset{{\cc@style\underline{\mskip8mu}}}}
\makeatother

\begin{document}

\title{Side-channel-free quantum key distribution with practical devices}

\author{Cong Jiang}
\affiliation{Jinan Institute of Quantum Technology, Jinan, Shandong 250101, P.~R.~China}
\affiliation{State Key Laboratory of Low Dimensional Quantum Physics, Department of Physics, Tsinghua University, Beijing 100084, P.~R.~China}

\author{Zong-Wen Yu}
\affiliation{Data Communication Science and Technology Research Institute, Beijing 100191, P.~R.~China}

\author{Xiao-Long Hu}
\affiliation{School of Physics, State Key Laboratory of Optoelectronic Materials and Technologies, Sun Yat-sen University, Guangzhou 510275, P.~R. China}

\author{Xiang-Bin Wang}\email{Corresponding author: xbwang@mail.tsinghua.edu.cn}
\affiliation{Jinan Institute of Quantum Technology, Jinan, Shandong 250101, P.~R.~China}
\affiliation{State Key Laboratory of Low Dimensional Quantum Physics, Department of Physics, Tsinghua University, Beijing 100084, P.~R.~China}
\affiliation{Shanghai Branch, CAS Center for Excellence and Synergetic Innovation Center in Quantum Information and Quantum Physics, University of Science and Technology of China, Shanghai 201315, P.~R.~China}
\affiliation{ Shenzhen Institute for Quantum Science and Engineering, and Physics Department, Southern University of Science and Technology, Shenzhen 518055, P.~R.~China}
\affiliation{Frontier Science Center for Quantum Information,
Beijing, P.~R.~China}

\begin{abstract}
Based on the idea that there is no side channel in the vacuum state, the side-channel-free quantum key distribution (SCFQKD) protocol was proposed, which is immune to all attacks in the source side-channel space and all attacks in the detectors. In the original SCFQKD protocol, an important assumption is that Alice and Bob can produce the perfect vacuum pulses. But due to the finite extinction ratio of the intensity modulators, the perfect vacuum pulse is impossible in practice. In this paper, we solve this problem and make the quantum key distribution side-channel secure with real source device which does not emit perfect vacuum pulses. Our conclusion only depends on the upper bounds of the intensities of the sources. No other assumptions such as stable sources and stable side channels are needed. The numerical results show that, comparing with the results of SCFQKD protocol with perfect vacuum sources, the key rates and secure distance are only slightly decreased if the upper bound of the intensity of the imperfect vacuum source is less than $10^{-8}$ which can be achieved in experiment by two-stage intensity modulator. We also show that the two-way classical communication can be used to the data post-processing of SCFQKD protocol to improve the key rate. Specially, the active odd-parity pairing method can improve the key rates in all distances by about two times and the secure distance by about 40 km. Give that the side channel security based on imperfect vacuum, this work makes it possible to realize side channel secure QKD with real devices.  
\end{abstract}


\maketitle
\section{Introduction}
Quantum key distribution can provide secure communication between two remote parties~\cite{bennett1984quantum,gisin2002quantum,xu2020secure,pirandola2020advances,scarani2009security,hwang2003quantum,wang2005beating,lo2005decoy,
lo2012measurement,braunstein2012side,wang2019practical}, no matter what the eavesdropper, Eve, does in the channel. However, the security of a practical QKD system can be broken due to the imperfection of the practical devices. In the source side, there  could be side channels which might leak extra information to Eve, such as basis dependent synchronization errors in the pulse emitting time or the frequency-spectrum difference for different encoding states or bases in the source sides. Besides, the detectors can be controlled by the eavesdropper through strong light attacks~\cite{lydersen2010hacking,gerhardt2011full,weier2011quantum}. Encoding the bits into sending or not-sending as used in the sending-or-not-sending (SNS) protocol~\cite{wang2018twin} of twin-field QKD~\cite{lu2018overcoming}, using the fact that there is no side channel in the vacuum state, the side-channel-free (SCF)QKD protocol was proposed~\cite{wang2019practical}. SCFQKD protocol~\cite{wang2019practical} is immune to all attacks in the side-channel space of sources, and by introducing a third party as a measurement station, it is also measurement device independent~\cite{lo2012measurement,braunstein2012side} immune to all attacks in the detectors. SCFQKD protocol only needs to know the upper bounds of the intensities of the non-vacuum sources and its secure distance can exceed 200 km even with $20\%$ misalignment error. Recently, SCFQKD protocol was experimentally demonstrated in 50 km fibers~\cite{zhang2022experimental}, which shows the potential of SCFQKD protocol in practical applications.     

Although the recent experiment~\cite{zhang2022experimental} has verified the the most impressive advantage of the promised long distance by SCFQKD protocol proposed in Ref.~\cite{wang2019practical}, the major problem in the original SCFQKD protocol requesting perfect vacuum source is still open. In this paper, we solve this open problem and make the QKD side-channel secure with real source device which does not emit perfect vacuum pulses. Our conclusion only depends on the upper bounds of the intensities of the sources. No other assumptions such as stable sources and stable side channels are needed. The channel security of our protocol allows whatever imperfect detection loophole and whatever side channel imperfection of emitted  photons, say, it guarantees a secure QKD provided that Eve has no access to devices inside Alice's and Bob's labs. Give that the side channel security based on imperfect vacuum, this work makes it possible to realize side channel secure QKD with real devices.

The paper is arranged as follows. We first introduce the procedure of SCFQKD protocol with real devices in Sec.~\ref{protocol}. We then show how to estimate the phase-flip error rate of a certain time window in Sec.~\ref{one_window}. With the conclusion in Sec.~\ref{one_window}, we further generalize the estimation method of the phase-flip error rate to the whole protocol and get the key rate formula. The numerical simulation results are shown in Sec.~\ref{simulation} where we also show that the two-way classical communication (TWCC) can be used to the data post-processing of SCFQKD protocol to improve the key rate. The article is ended with some conclusion remarks.

\section{The protocol}\label{protocol}
For the time window $i$, Alice (Bob) randomly chooses the weak source, i.e., the imperfect vacuum source $o_A$ ($o_B$), or the strong source $x_A$ ($x_B$) with probabilities $p_0$ and $p_x=1-p_0$ respectively. If the weak source $o_A$ ($o_B$) is chosen, a weak coherent state (WCS) pulse with intensity $\nu_A^i$ ($\nu_B^{i}$) is prepared, and Alice (Bob) takes it as bit $0$ ($1$). If the strong source $x_A$ ($x_B$) is chosen, a WCS pulse with intensity $\mu_A^i$ ($\mu_B^{i}$) is prepared, and Alice (Bob) takes it as bit $1$ ($0$). Alice and Bob send the prepared pulses to a untrusted third party, Charlie, who is assumed to first compensate the phase difference of the received pulse pair and then preform the interference measurement. Charlie would publicly announce the measurement results to Alice and Bob. If only one detector clicks, Alice and Bob would take the $i$-th window as an effective window, and this event is also called an effective event whose corresponding bit is called an effective bit.

After Alice and Bob repeat the above process for $N$ times and Charlie announces all the measurement results, \emph{they} perform the data post-processing. For each time window, Alice randomly decides whether it is a test window which is used for decoy analysis with probability $r$, or a key generation window which is used for the final key distillation with probability $1-r$. For the effective test windows, Alice and Bob publicly announce the sources they used in each time windows. For the effective key generation windows, the corresponding bits are used to distil the final keys.  

For a time window, if only one of Alice and Bob decides to send out a pulse from strong sources, it is a $\tilde{Z}$ window. For a time window, if both Alice and Bob decide to send out a pulse from strong sources (weak sources), it is a $\mathcal{B}$ ($\mathcal{O}$) window. 

The corresponding effective bits of the effective events of $\tilde{Z}$ key generation windows are untagged bits. The $\tilde{Z}$ key generation window means it is a $\tilde{Z}$ window and chosen for key generation. Through decoy state analysis, we can get the upper bound of the phase-flip error rate of those untagged bits, $\overline{e}^{ph}$. The key rate formula is
\begin{equation}\label{key_rate}
R=\frac{1}{N}\{n_{u}[1-H(\overline{e}^{ph})]-fn_tH(E_K)\},
\end{equation}       
where $H(x)=-x\log_2x-(1-x)\log_2(1-x)$ is the entropy function; $n_{{u}}$ is the number of untagged bits; $n_t$ is the number of corresponding bits of effective key generation windows; $f$ is the correction efficiency factor; $E_K$ is the bit-flip error rate of the effective bits from the key generation windows.

In what follows we shall study how to calculate $n_u$ and $\overline{e}^{ph}$ by observed values and we result in Eqs.~(\ref{nod}-\ref{eph}).

\section{The phase-flip error rate of a certain time window}\label{one_window}
We first consider the phase-flip error rate of a certain time window. For simplicity, we omit the superscript $i$ of $\nu_A^i,\nu_B^{i},\mu_A^i,\mu_B^{i}$ and all other physical quantities and states appeared in this section. 

In a real experiment, instead of simply living in the operational space (Fock space), the sent out pulses actually live in the whole space including all side channel spaces such as the frequency, the polarization, the spatial angular momentum and so on. Yet the vacuum state has no side-channel space and therefore we only need to consider the side-channel space for the non vacuum parts. The states can be decomposed in two parts, vacuum and non vacuum. Explicitly, if Alice (Bob) chooses the weak source, she (he) actually prepares the state:
\begin{equation}\label{real_state1}
\begin{split}
&\ket{\alpha_A^0}=e^{-\nu_A/2}\ket{0}+\sqrt{1-e^{-\nu_A}}\ket{\psi_A},\\
&\ket{\alpha_B^0}=e^{-\nu_B/2}\ket{0}+\sqrt{1-e^{-\nu_B}}\ket{\psi_B}.\\
\end{split}
\end{equation}
If Alice (Bob) chooses the strong source, she (he) actually prepares the state:
\begin{equation}\label{real_state2}
\begin{split}
&\ket{\alpha_A}=e^{-\mu_A/2}\ket{0}+\sqrt{1-e^{-\mu_A}}\ket{\phi_A},\\
&\ket{\alpha_B}=e^{-\mu_B/2}\ket{0}+\sqrt{1-e^{-\mu_B}}\ket{\phi_B}.
\end{split}
\end{equation}
Here $\ket{0}$ is the vacuum state and $\ket{\psi_A},\ket{\psi_B},\ket{\phi_A},\ket{\phi_B}$ are the corresponding non vacuum parts of each states. Obviously, we have 
\begin{equation}
\iprod{0}{\psi_A}=\iprod{0}{\psi_B}=\iprod{0}{\phi_A}=\iprod{0}{\phi_B}=0,
\end{equation}
which would be used in the calculation of phase-flip error rate.
 
As shall be shown, our protocol does not request any specific photon number distribution of its sources, it only needs a linear superposition of vacuum and non vacuum for the source state in whole space.
 
In this certain time window, if only one of Alice and Bob chooses the strong source, it is a $\tilde{Z}$ window. To prove the security, we consider the virtual protocol where Alice and Bob preshare
\begin{equation}
\ket{\Psi}=\frac{1}{\sqrt{2}}(\ket{\alpha_A^0,\alpha_B}\otimes \ket{01}_{\mathcal{I}}+\ket{\alpha_A,\alpha_B^0}\otimes \ket{10}_{\mathcal{I}}).
\end{equation}
Also, we have
\begin{equation}
\ket{\Psi}=\frac{1}{2}(\mathcal{N}_+\ket{\chi^+}\otimes \ket{\Phi^0}_{\mathcal{I}}+\mathcal{N}_-\ket{\chi^-}\otimes \ket{\Phi^1}_{\mathcal{I}}),
\end{equation}
where 
\begin{equation}
\ket{\Phi^0}_{\mathcal{I}}=\frac{1}{\sqrt{2}}(\ket{01}_{\mathcal{I}}+\ket{10}_{\mathcal{I}}),\quad \ket{\Phi^1}_{\mathcal{I}}=\frac{1}{\sqrt{2}}(\ket{01}_{\mathcal{I}}-\ket{10}_{\mathcal{I}}),
\end{equation}
and
\begin{equation}
\ket{\chi^+}=\frac{1}{\mathcal{N}_+}(\ket{\alpha_A^0,\alpha_B}+\ket{\alpha_A,\alpha_B^0}), \quad \ket{\chi^-}=\frac{1}{\mathcal{N}_-}(\ket{\alpha_A^0,\alpha_B}-\ket{\alpha_A,\alpha_B^0}),
\end{equation}
where $\mathcal{N}_+$ and $\mathcal{N}_-$ are normalization coefficients.

Here $\ket{01}_{\mathcal{I}}$ and $\ket{10}_{\mathcal{I}}$ are local states that are stored in Alice's and Bob's labs. If Alice and Bob decide to measure their local states in $Z$ basis, i.e., $\{\ket{01}_{\mathcal{I}}, \ket{10}_{\mathcal{I}}\}$ before \emph{they} send out the pulse pair, it is equivalent to a protocol where Alice and Bob randomly send out a pulse pair in state $\ket{\alpha_A^0,\alpha_B}$ or $\ket{\alpha_A,\alpha_B^0}$ with $50\%$ probability. If Alice and Bob decide to measure their local states in $X$ basis, i.e., $\{\ket{\Phi^0}_{\mathcal{I}}, \ket{\Phi^1}_{\mathcal{I}}\}$ before \emph{they} send out the pulse pair, it is equivalent to a protocol where Alice and Bob randomly send out a pulse pair in state $\ket{\chi^+}$ or $\ket{\chi^-}$ with probabilities $\frac{\mathcal{N}_+^2}{4}$ and $\frac{\mathcal{N}_-^2}{4}$ respectively.  

In this protocol, a phase error occurs in either of the following two kinds of effective windows: 1) the effective window while Alice and Bob send out a pulse pair in state $\ket{\chi^+}$, i.e., the measurement result of \emph{their} local state is $\ket{\Phi^0}_{\mathcal{I}}$, and Charlie announces the right detector clicking; 2) the effective window while Alice and Bob send out a pulse pair in state $\ket{\chi^-}$, i.e., the measurement result of \emph{their} local state is $\ket{\Phi^1}_{\mathcal{I}}$, and Charlie announces the left detector clicking.

We denote $S_{\zeta}^d$ as the probability that Charlie announces an effective event with detector $d$ clicking in a time window when \emph{they} have sent out state from source $\zeta$. Here $d\in\{L,R\}$ and $\zeta\in\{\mathcal{O},\mathcal{B},\tilde{Z}\}$; $L$ represents the left detector and $R$ represents the right detector. We denote $S_{X_+}^d$ ($S_{X_-}^d$) as the probability that Charlie announces an effective event with detector $d$ clicking in a time window when \emph{they} have sent out state $\ket{\chi^+}$ ($\ket{\chi^-}$).

With all those definitions, we can express the probability that Alice and Bob detect a phase error in the $\tilde{Z}$ window, $T_X$, as the following form
\begin{equation}
T_X=\frac{\mathcal{N}_+^2}{4} S_{X_+}^R+\frac{\mathcal{N}_-^2}{4} S_{X_-}^L=\frac{\mathcal{N}_+^2}{4} (S_{X_+}^R- S_{X_+}^L)+S_{\tilde{Z}}^L.
\end{equation}
Here we use the fact that density matrices of the sent out pulse pairs are the same when Alice and Bob measure their local states in $X$ basis and $Z$ basis, and thus $\frac{\mathcal{N}_+^2}{4} S_{X_+}^L+\frac{\mathcal{N}_-^2}{4} S_{X_-}^L=S_{\tilde{Z}}^L$.

We also have the phase-flip error rate in the $\tilde{Z}$ window
\begin{equation}\label{eph_one}
e^{ph}=\frac{T_X}{S_{\tilde{Z}}}=\frac{\frac{\mathcal{N}_+^2}{4} (S_{X_+}^R- S_{X_+}^L)+S_{\tilde{Z}}^L}{S_{\tilde{Z}}},
\end{equation} 
where $S_{\tilde{Z}}=S_{\tilde{Z}}^L+S_{\tilde{Z}}^R$.

As shown in the Appendix~\ref{cal1}, we have the upper bound of $S_{X_+}^R$ and the lower bound of $S_{X_-}^L$
\begin{align}
\label{sx+}&{S}_{X_+}^R\le \frac{1}{\mathcal{N}_+^2}\left(c_0^2 S_{\mathcal{O}}^R+c_1^2 S_{\mathcal{B}}^R+c_2^2+2c_0c_1\sqrt{S_{\mathcal{O}}^RS_{\mathcal{B}}^R}+2c_0c_2\sqrt{S_{\mathcal{O}}^R}+2c_1c_2\sqrt{S_{\mathcal{B}}^R}\right),\\
\label{sx-}&{S}_{X_+}^L\ge  \frac{1}{\mathcal{N}_+^2}\left(c_0^2 S_{\mathcal{O}}^L+c_1^2 S_{\mathcal{B}}^L-2c_0c_1\sqrt{S_{\mathcal{O}}^LS_{\mathcal{B}}^L}-2c_0c_2\sqrt{S_{\mathcal{O}}^L}-2c_1c_2\sqrt{S_{\mathcal{B}}^L}\right),
\end{align}
where $c_0,c_1,c_2$ are real positive values, $c_0c_1=1$ and 
\begin{equation}\label{c2i}
c_2^2\le \left(c_0+c_1-2e^{-\nu_A/2-\mu_A/2}+2\sqrt{1-e^{-\nu_A}}\sqrt{1-e^{-\mu_A}}\right)\left(c_0+c_1-2e^{-\nu_B/2-\mu_B/2}+2\sqrt{1-e^{-\nu_B}}\sqrt{1-e^{-\mu_B}}\right).
\end{equation}
With formulas above, we can get the upper bound of $T_X$.

\section{The phase-flip error rate in the whole protocol}\label{all_window}
In Sec.~\ref{one_window}, we get the phase-error rate of a certain $\tilde{Z}$ window. But in practice, the sources are usually unstable in the whole spaces, which means the intensities of the sources and the actual states in different time windows might be different. Thus we can not directly take Eq.~\eqref{eph_one} as the formula of the upper bound of the phase-flip error rate in the whole protocol. However, Eq.~\eqref{eph_one} holds for any certain $\tilde{Z}$ window, provided that we replace all values including the intensities $\mu$, the probabilities $S_{\zeta}^d$, and $T_X$, $c_0,c_1,c_2$ by the corresponding values in this certain $\tilde{Z}$ window.

Recall that $T_X^i$ is the probability that a phase error occurs if the $i$-th window is a $\tilde{Z}$ window, we have
\begin{equation}
n^{ph}=\sum_{i=1}^N 2p_0p_x(1-r) T_X^i,
\end{equation}
where $n^{ph}$ is the number of phase errors in the $\tilde{Z}$ key generation windows of the whole protocol.

Eqs. (\ref{sx+}-\ref{c2i}) always hold provided that $c_0^ic_1^i=1$. Thus we take the same value of $c_0^i$ and $c_1^i$ for all time windows and denote by $c_0,c_1$ respectively. The valve of $c_2^i$ is upper bounded by Eq.~\eqref{c2i}. Further more, we have
\begin{equation}\label{c2bar}
(c_2^i)^2\le \overline{c}_2^2=\left(c_0+c_1-2e^{-\nu_A^U/2-\mu_A^U/2}+2\sqrt{1-e^{-\nu_A^U}}\sqrt{1-e^{-\mu_A^U}}\right)\left(c_0+c_1-2e^{-\nu_B^{ U}/2-\mu_B^U/2}+2\sqrt{1-e^{-\nu_B^{ U}}}\sqrt{1-e^{-\mu_B^U}}\right),
\end{equation}
where $\nu_A^U,\mu_A^U,\nu_B^{ U},\mu_B^U$ are the upper bounds of $\nu_A^i,\mu_A^i,\nu_B^{i},\mu_B^i$ respectively and we assume those bounds are known values in the protocol. $\overline{c}_2$ is the upper bound of $c_2^i$ for all time windows. We have
\begin{equation}
\begin{split}
n^{ph}=&\sum_{i=1}^N 2p_0p_x(1-r) T_X^i\\
\le & \sum_{i=1}^N \frac{1}{2}p_0p_x(1-r)\left[c_0^2(S_{\mathcal{O}}^{i,R}-S_{\mathcal{O}}^{i,L})+c_1^2(S_{\mathcal{B}}^{i,R}-S_{\mathcal{B}}^{i,L})+\overline{c}_2^2+2c_0c_1\left(\sqrt{S_{\mathcal{O}}^{i,R}S_{\mathcal{B}}^{i,R}}+\sqrt{S_{\mathcal{O}}^{i,L}S_{\mathcal{B}}^{i,L}}\right)  \right.\\
&\left.+2c_0\overline{c}_2\left(\sqrt{S_{\mathcal{O}}^{i,R}}+\sqrt{S_{\mathcal{O}}^{i,L}}\right)+2c_1\overline{c}_2\left(\sqrt{S_{\mathcal{B}}^{i,R}}+\sqrt{S_{\mathcal{B}}^{i,L}}\right)\right]+\sum_{i=1}^N 2p_0p_x(1-r) S_{\tilde{Z}}^{i,L}\\
\le& \frac{1}{2}p_0p_x(1-r)\left[c_0^2\sum_{i=1}^N(S_{\mathcal{O}}^{i,R}-S_{\mathcal{O}}^{i,L})+c_1^2\sum_{i=1}^N(S_{\mathcal{B}}^{i,R}-S_{\mathcal{B}}^{i,L})+\overline{c}_2^2+2c_0c_1\left(\sqrt{\sum_{i=1}^NS_{\mathcal{O}}^{i,R}\sum_{i=1}^NS_{\mathcal{B}}^{i,R}}+\sqrt{\sum_{i=1}^NS_{\mathcal{O}}^{i,L}\sum_{i=1}^NS_{\mathcal{B}}^{i,L}}\right)  \right.\\
&\left.+2c_0\overline{c}_2\left(\sqrt{N\sum_{i=1}^NS_{\mathcal{O}}^{i,R}}+\sqrt{N\sum_{i=1}^NS_{\mathcal{O}}^{i,L}}\right)+2c_1\overline{c}_2\left(\sqrt{N\sum_{i=1}^NS_{\mathcal{B}}^{i,R}}+\sqrt{N\sum_{i=1}^NS_{\mathcal{B}}^{i,L}}\right)\right]+\sum_{i=1}^N 2p_0p_x(1-r) S_{\tilde{Z}}^{i,L}.
\end{split}
\end{equation}
Here we have used the Cauchy inequality in the second inequality
\begin{equation}
\left(\sum_{i=1}^Na_ib_i \right)^2\le \sum_{i=1}^Na_i^2 b_i^2\quad a_i,b_i\in \mathbb{R}. 
\end{equation}

Denote $n_{\zeta}^d$ as the number of observed effective events caused by the detector $d$ in the $\zeta$-test-windows (those $\zeta$ windows chosen for test) where $d\in\{L,R\}$ and $\zeta\in\{\mathcal{O},\mathcal{B},\tilde{Z}\}$. We have
\begin{align}
\label{nod}n_{\mathcal{O}}^d=\sum_{i=1}^N p_0^2 r S_{\mathcal{O}}^{i,d},\quad n_{\mathcal{B}}^d=\sum_{i=1}^N p_x^2 r S_{\mathcal{B}}^{i,d},\quad n_{\tilde{Z}}^d=\sum_{i=1}^N 2p_0p_x r S_{\tilde{Z}}^{i,d}.
\end{align}
We define
\begin{equation}
S_{\mathcal{O},A}^d=\frac{n_{\mathcal{O}}^d}{Np_0^2 r},\quad S_{\mathcal{B},A}^d=\frac{n_{\mathcal{B}}^d}{Np_x^2 r},\quad S_{\tilde{Z},A}^d=\frac{n_{\tilde{Z}}^d}{2Np_0p_x r}.
\end{equation}
   
With those observed values, we have
\begin{equation}\label{nph}
\begin{split}
n^{ph}\le& \overline{n}^{ph}\\
= & \frac{1}{2}p_0p_x(1-r)N\left[c_0^2(S_{\mathcal{O},A}^R-S_{\mathcal{O},A}^L)+c_1^2(S_{\mathcal{B},A}^R-S_{\mathcal{B},A}^L)+\overline{c}_2^2+2c_0c_1\left(\sqrt{S_{\mathcal{O},A}^RS_{\mathcal{B},A}^R}+\sqrt{S_{\mathcal{O},A}^LS_{\mathcal{B},A}^L}\right)  \right.\\
&\left.+2c_0\overline{c}_2\left(\sqrt{S_{\mathcal{O},A}^R}+\sqrt{S_{\mathcal{O},A}^L}\right)+2c_1\overline{c}_2\left(\sqrt{S_{\mathcal{B},A}^R}+\sqrt{S_{\mathcal{B},A}^L}\right)+4S_{\tilde{Z},A}^L\right].
\end{split}
\end{equation}
And the number of untagged bits $n_u$ satisfies
\begin{equation}\label{nu}
n_u=\sum_{i=1}^N 2p_0p_x(1-r)(S_{\tilde{Z}}^{i,L}+S_{\tilde{Z}}^{i,R})=2p_0p_x(1-r)N(S_{\tilde{Z},A}^L+S_{\tilde{Z},A}^R).
\end{equation}
Then, we get the upper bound of the phase-flip error rate of the untagged bits in the key generation windows
\begin{equation}\label{eph}
\overline{e}^{ph}=\frac{\overline{n}^{ph}}{n_u}.
\end{equation}
With Eqs.~(\ref{nod}-\ref{eph}), we can calculate the key rate by Eq.~\eqref{key_rate}.

{\bf{Remark:}} Although we have used model of WCS sources in the calculation above, it's quite obvious that our method here can apply to any type of source since we can always express the states of any sources into the linear superposition of vacuum part and non vacuum part:
\begin{equation}
\ket{\mathcal{A}}=\sqrt{a_0}\ket{0}+\sqrt{1-a_0}\ket{{\rm non-vacuum}},
\end{equation}
where $a_0$ is the probability of vacuum part of the state and $\ket{{\rm non-vacuum}}$ is a whole space non vacuum state. As shown in Eq.~\eqref{c2bar}, our method only depends on $e^{-\nu_A^U},e^{-\nu_B^U},e^{-\mu_A^U},e^{-\mu_B^U}$, i.e., the lower bounds of the probabilities of vacuum state, thus Eq.~\eqref{c2bar} holds for any sources provided that we replace those lower bounds by the corresponding lower bound of $a_0$. We can get the lower bound of $a_0$ by partially characterizing states in Fock space. Specially, for the WCS sources, we can get the lower bound of $a_0$ by measuring the upper bound of the intensities $\nu$ or $\mu$.  

\section{Numerical simulation}\label{simulation}
We shall consider the symmetry case here. In the symmetry case, the distance from Alice to Charlie is the same as the distance from Bob to Charlie. And Charlie's two detectors are assumed to have the same properties such as the dark counting rate and the detection efficiency. Without loss of generality, we assume the source parameters of Alice and Bob are the same, i.e., $\nu_A^U=\nu_B^{ U}=\nu$ and $\mu_A^U=\mu_B^U=\mu$. In the calculation of key rate, $c_0,c_1$ can be taken as any positive real values provided that $c_0c_1=1$, and we can optimize $c_0,c_1$ to achieve the highest key rate. For simplicity, we set $c_0=e^{\nu/2-\mu/2},c_1=e^{\mu/2-\nu/2}$. The experiment parameters are listed in Table \ref{exproperty}. In the numerical simulation, $\nu$ is a fixed value and the other source parameters including $p_0,p_x,\mu$ are optimized. Since the asymptotic case is considered here, we ignore the influence to the key rate of $r$, i.e., we take $r\sim 0$.

\begin{table}[h]
\centering
\begin{tabular}{ccccc}
\hline
$p_d$ & $E_d$ &$\eta_d$ & $f$ & $\alpha_f$\\
\hline
$1.0\times10^{-9}$& $4\%$ & $60.0\%$ & $1.1$ & $0.2$\\ 
\hline
\end{tabular}

\caption{List of experimental parameters used in numerical simulations. Here $p_d$ is the dark counting rate per pulse of Charlie's detectors; $\eta_d$ is the detection efficiency of Charlie's detectors; $E_d$ is the misalignment error; $f$ is the error correction inefficiency; $\alpha_f$ is the fiber loss coefficient (dB/km).}\label{exproperty}
\end{table}

\begin{figure}[h]
\centering
\includegraphics[width=8cm]{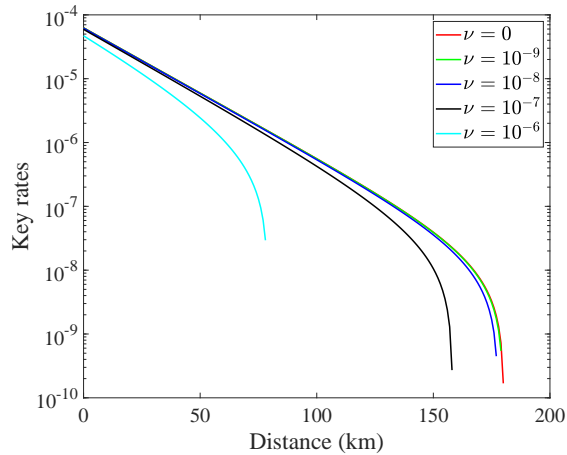}
\caption{The comparison of the key rates of SCFQKD protocol under different $\nu$. The experiment parameters here are listed in Table \ref{exproperty}.}\label{org1}
\end{figure}

\begin{figure}[h]
\centering
\includegraphics[width=8cm]{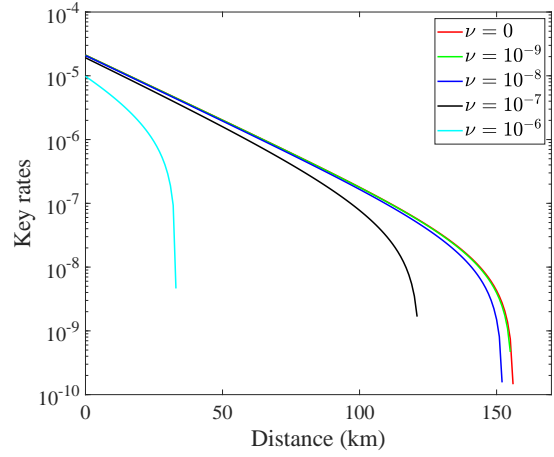}
\caption{The comparison of the key rates of SCFQKD protocol under different $\nu$. We set $E_d=10\%$. The other experiment parameters are listed in Table \ref{exproperty}.}\label{org2}
\end{figure}

Figures~\ref{org1} and \ref{org2} are the key rates of SCFQKD protocol under different $\nu$. The experiment parameters listed in Table \ref{exproperty} are used here, except for we set $E_d=10\%$ in Figure~\ref{org2}. By setting $\nu=0$, the key rate formulas in Eqs.~(\ref{key_rate},\ref{nod}-\ref{eph}) are the same with those of the original SCFQKD protocol~\cite{wang2019practical}. Thus lines `$\nu=0$' in Figures~\ref{org1} and \ref{org2} are the results of SCFQKD protocol with perfect vacuum sources, i.e., the original SCFQKD protocol. Results in Figures~\ref{org1} and \ref{org2} show that the imperfect vacuum sources, i.e., the weak sources have little affect on the key rates if the upper bound of the intensities of the the imperfect vacuum sources are lower than $10^{-8}$. But when the upper bound of the intensities of the the imperfect vacuum sources is as large as $10^{-6}$, the key rates and secure distances are drastically decreased compare with those of the original SCFQKD protocol. In experiments, the intensity of the imperfect vacuum sources can be controlled in the level of $10^{-8}$ by two-stage intensity modulator~\cite{zhang2022experimental}, thus we can expect little affect on the key rates in experiment due to the imperfect vacuum sources.       

Since there are no errors in the untagged bits of SCFQKD protocol with or without perfect vacuum sources. We can directly applied the TWCC methods~\cite{xu2020sending,jiang2021composable} including the standard TWCC method and the AOPP method to improve the key rates and secure distance of the SCFQKD protocol. The calculation methods are shown in Appendix~\ref{twcc_method}.

Figures~\ref{ed04} and \ref{ed10} are the comparison of the key rates of SCFQKD protocol with or without TWCC. The `Original' lines are the results calculated by Eq.~\eqref{key_rate}. The `Standard TWCC' lines are the results calculated by Eq.~\eqref{key_rate2}. The `AOPP' lines are the results calculated by Eq.~\eqref{key_rate3}. We set $E_d=10\%, \nu=0$ in Figure~\ref{ed04}, and  $\nu=10^{-8}$ in Figure~\ref{ed10}. The other experiment parameters are listed in Table \ref{exproperty}. Results in Figures~\ref{ed04} and \ref{ed10} show that both the standard TWCC method and the AOPP method can improve the secure distance by about 40 km. The AOPP method can improve the key rates in all distances by about two times, while the standard TWCC method can only improve the key rates at long distance.  

\begin{figure}[h]
\centering
\includegraphics[width=8cm]{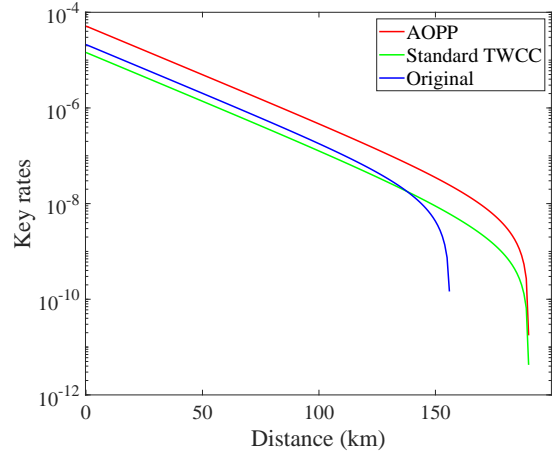}
\caption{The comparison of the key rates of SCFQKD protocol with or without TWCC. Here we set $E_d=10\%,\nu=0$. The other experiment parameters are listed in Table \ref{exproperty}.}\label{ed04}
\end{figure}

\begin{figure}[h]
\centering
\includegraphics[width=8cm]{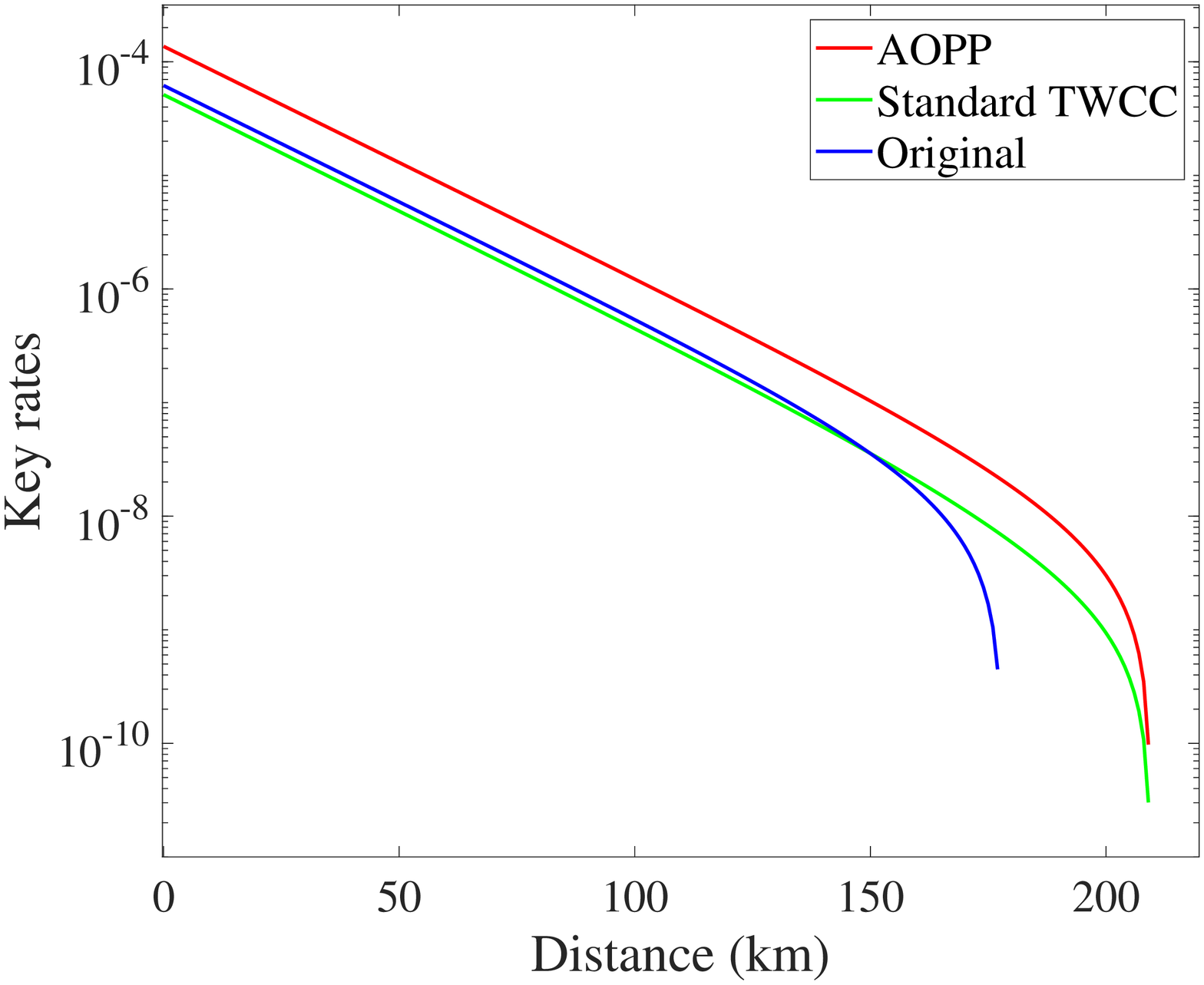}
\caption{The comparison of the key rates of SCFQKD protocol with or without TWCC. Here we set $\nu=10^{-8}$. The other experiment parameters are listed in Table \ref{exproperty}.}\label{ed10}
\end{figure}

\section{Conclusion}\label{conclusion}
In this paper, we make SCFQKD protocol side-channel secure with real source device which does not emit perfect vacuum pulses. Our conclusion only depends on the upper bounds of the intensities of the sources and no other assumptions are needed. The numerical simulation shows that the key rates and secure distance are only slightly decreased if the upper bound of the intensities of the imperfect vacuum sources are less than $10^{-8}$ which can be achieved in experiment by two-stage intensity modulator~\cite{zhang2022experimental}. We also show that the TWCC methods including the standard TWCC method and the AOPP method can be directly applied to SCFQKD protocol to improve the key rates and secure distance. Our numerical simulation results show that AOPP method can improve the key rates in all distances by about two times and improve the secure distance by about 40 km. Give that the side channel security based on imperfect vacuum, this work makes it possible to realize side channel secure QKD with real devices. Our protocol can also apply to efficient quantum digital signature by taking the post data processing method such as~\cite{amiri2016secure,qin2022quantum}. This will be reported elsewhere.

\section{Acknowledgement} 
We acknowledge the financial support in part by Ministration of Science and Technology of China through The National Key Research and Development Program of China Grant No. 2020YFA0309701; National Natural Science Foundation of China Grant Nos. 12174215, 12104184, 11974204 and 12147107; Shandong Provincial Natural Science Foundation Grant No. ZR2021LLZ007; Key R$\&$D Plan of Shandong Province Grant Nos. 2021ZDPT01; Open Research Fund Program of the State Key Laboratory of Low-Dimensional Quantum Physics Grant No. KF202110. 

\appendix
\section{The calculation method of the upper and lower bounds of $S_{X_+}^R$ and $S_{X_+}^L$}\label{cal1}
For a certain $\tilde{Z}$ window, we have
\begin{equation}\label{chi+}
\ket{\chi^+}=\frac{c_0\ket{\alpha_A^0,\alpha_B^0}+c_1\ket{\alpha_A,\alpha_B}+c_2\ket{\phi_2}}{\mathcal{N_+}},
\end{equation}
where 
\begin{equation}
c_2\ket{\phi_2}=\ket{\alpha_A^0,\alpha_B}+\ket{\alpha_A,\alpha_B^0}-c_0\ket{\alpha_A^0,\alpha_B^0}-c_1\ket{\alpha_A,\alpha_B}.
\end{equation}
Without loss of generality, we assume $c_0,c_1,c_2$ are real positive values. In principle, we can determine the values of $c_0$ and $c_1$ as we want and $c_2,\ket{\phi_2}$ are determined by $c_0,c_1$. For the convenience of the later calculation, we take $c_0c_1=1$.

Denote $\iprod{\psi_A}{\phi_A}=\beta_A$ and $\iprod{\psi_B}{\phi_B}=\beta_B$. Using the normalization condition, we have
\begin{equation}
c_2^2=2+c_0^2+c_1^2+(\gamma_A\gamma_B^*+\gamma_A^*\gamma_B)-(c_0+c_1)(\gamma_A^*+\gamma_B^*+\gamma_A+\gamma_B)+c_0c_1(\gamma_A^*\gamma_B^*+\gamma_A\gamma_B),
\end{equation}
where
\begin{align}
\gamma_A=e^{-\nu_A/2-\mu_A/2}+\sqrt{1-e^{-\nu_A}}\sqrt{1-e^{-\mu_A}}\beta_A,\\
\gamma_B=e^{-\nu_B/2-\mu_B/2}+\sqrt{1-e^{-\nu_B}}\sqrt{1-e^{-\mu_B}}\beta_B.
\end{align}

With the condition $c_0c_1=1$, we have
\begin{equation}
c_2^2=(c_0+c_1-\gamma_A-\gamma_A^*)(c_0+c_1-\gamma_B-\gamma_B^*).
\end{equation}

It is easy to check that the worst case of the phase-flip error rate is achieved when $\beta_A=\beta_B=-1$. And we have
\begin{equation}
c_2^2\le \left(c_0+c_1-2e^{-\nu_A/2-\mu_A/2}+2\sqrt{1-e^{-\nu_A}}\sqrt{1-e^{-\mu_A}}\right)\left(c_0+c_1-2e^{-\nu_B/2-\mu_B/2}+2\sqrt{1-e^{-\nu_B}}\sqrt{1-e^{-\mu_B}}\right).
\end{equation}

Finally, apply the input-output theory proposed in Ref.~\cite{wang2019practical}, we can get the upper and lower bounds of $S_{X_+}^R$ and $S_{X_+}^L$ shown in Eqs.~(\ref{sx+},\ref{sx-}). To ensure the completeness of the article, this theory is briefly introduced in Appendix~\ref{input}.

\section{The input-output theory}\label{input}
The key idea of the input-output theory is that in a certain time window, we can regard Charlie uses the same measurement process to measure the received quantum state no matter what the quantum state is. This theory is proposed in Ref.~\cite{wang2019practical}, here we just simply introduce its content.

Suppose at the beginning of a certain time window, Alice and Bob send out a pulse pairs in state $\ket{\psi}$. Charlie, who is assumed to control the channel and measurement station, then combines this state with his ancillary state $\ket{\kappa}$. Charlie's instrument state $\mathcal L$ is included in the ancillary state $|\kappa\rangle$. The initial state is
\begin{equation}
 |\Psi_{ini}\rangle = |\psi\rangle\otimes|\kappa\rangle.
\end{equation}

  At time $t$, Charlie observes his instrument $\mathcal L$ to see the result. His instrument $\mathcal L$ is observed by Alice and she can find  result from $\{l_i\}$ accompanied with its eigenstate $| l_i\rangle$ then.
  Most generally, after state $|\psi\rangle$ is sent to Charlie, Charlie's initial state $|\Psi_{ini}\rangle = |\psi\rangle\otimes|\kappa\rangle $ will evolve with time under a quantum process. Here we  assume a unitary quantum process $\mathcal U$.  Even though Charlie presents a non-unitary quantum process, it can be represented
  by a unitary process through adding more ancillary states. So, given the general ancillary state $|\kappa\rangle$, we can simply assume a unitary quantum process for Charlie.  At time $t$, the state is now
 \begin{equation}
 |\Psi(t)\rangle = \mathcal U(t)|\Psi_{ini}\rangle=\mathcal U(t)(|\psi\rangle\otimes|\kappa\rangle)
 \end{equation}
 In general, the state at time $t$ can be written in a bipartite form of another two subspaces, one is the instrument space $\mathcal L$ and the other is the remaining part of the space, subspace $\bar {\mathcal L}$.
 Given the initial input state $|\psi\rangle$ to Charlie, the probability that he observes the result $l_1$ at time $t$ is
 \begin{equation}\label{p0}
 S_{\ket{\psi}}^{l_1}=\langle l_1|\tr_{\bar{\mathcal L}}\left(|\Psi(t)\rangle\langle\Psi(t)|\right)|l_1\rangle
 \end{equation}
We will omit $(t)$ in the following formulas.
 Suppose the space $\bar {\mathcal L}$ is spanned by basis states $\{g_k\}$, we can rewrite Eq.(\ref{p0}) by
 \begin{align}\label{basic}
 & S_{\ket{\psi}}^{l_1} =\sum_k |\langle \gamma_k^{(l_1)}|\Psi\rangle|^2
 \end{align}
 where $|\gamma_k^{(l_1)}\rangle=|g_k\rangle|l_1\rangle$.

Suppose state $|\phi\rangle$ has the form of
\begin{equation}\label{equ:phi}
    \ket{\phi} = \xi_0 \ket{\phi_0} + \xi_1 \ket{\phi_1} + \xi_2 \ket{\phi_2},
\end{equation}
Without loss of generality, we assume $\xi_0,\xi_1,\xi_2$ are real positive values. With Eq.~\eqref{basic}, we have
\begin{align}
&S_{\ket{\phi}}^{l_1} =\sum_k |\langle \gamma_k^{(l_1)}|\phi\rangle|^2,\\
&S_{\ket{\phi_0}}^{l_1} =\sum_k |\langle \gamma_k^{(l_1)}|\phi_0\rangle|^2,\\
\label{eqf}&S_{\ket{\phi_1}}^{l_1} =\sum_k |\langle \gamma_k^{(l_1)}|\phi_1\rangle|^2,\\
\end{align}
where $S_{\ket{\tau}}^{l_1}$ is the probability that Charlie observes the result $l_1$ at time $t$ if Alice and Bob send out a pulse in state $\ket{\tau}$ in a certain  time window for $\tau=\phi,\phi_0,\phi_1$.

With Eqs.~(\ref{equ:phi}-\ref{eqf}), we have
\begin{equation}\label{equ:SphiU}
\begin{split}
S_{\ket{\phi}}^{l_1} \le \xi_0^2 S_{\ket{\phi_0}}^{l_1} + \xi_1^2 S_{\ket{\phi_1}}^{l_1} + \xi_2^2+ 2\xi_0 \xi_1 \sqrt{S_{\ket{\phi_0}}^{l_1} S_{\ket{\phi_1}}^{l_1}} + 2\xi_0 \xi_2 \sqrt{S_{\ket{\phi_0}}^{l_1}}  + 2\xi_1 \xi_2 \sqrt{S_{\ket{\phi_1}}^{l_1}}
\end{split}
\end{equation}
and
\begin{equation}\label{equ:SphiL}
\begin{split}
S_{\ket{\phi}}^{l_1} \ge \xi_0^2 S_{\ket{\phi_0}}^{l_1} + \xi_1^2 S_{\ket{\phi_1}}^{l_1} -\left( 2\xi_0 \xi_1 \sqrt{S_{\ket{\phi_0}}^{l_1} S_{\ket{\phi_1}}^{l_1}} + 2\xi_0 \xi_2 \sqrt{S_{\ket{\phi_0}}^{l_1}}  + 2\xi_1 \xi_2 \sqrt{S_{\ket{\phi_1}}^{l_1}}\right)
\end{split}
\end{equation}

\section{The TWCC methods}\label{twcc_method}
Before Alice and Bob perform the error correction, \emph{they} can first perform the TWCC methods to reduce the bit-flip error rate in the raw keys. Both the standard TWCC method and the AOPP method can be applied to the SCFQKD protocol~\cite{xu2020sending}. And the iteration formulas of the lower bound of the untagged bits and the upper bound of the phase-flip error rate after TWCC are also holds here~\cite{xu2020sending}.

To perform the standard TWCC, Bob first randomly pairs his bits two by two and then announces all the paired sequences to Alice through the public channel. Then Alice and Bob compare the parity of these bit pairs, \emph{they} keep one bit from the bit pairs with same parities and discard the rest. The survived bits form a new bit string and would be performed the error correction and privacy amplification to distil the final keys according to the following key rate formulas
\begin{equation}\label{key_rate2}
R^\prime=\frac{1}{N}\{n_{u}^{twcc}[1-H(\overline{e}_{ph}^{twcc})]-f[n_{t1}H(E_{1})+n_{t2}H(E_2)+n_{t3}H(E_3)]\}.
\end{equation}  
Here $n_{u}^{twcc}$ is the number of untagged bits after TWCC and
\begin{equation}
n_{u}^{twcc}=\frac{n_{u}^2}{2n_t}.
\end{equation}
$\overline{e}_{ph}^{twcc}$ is the upper bound of the phase-flip error rate after standard TWCC and
\begin{equation}
\overline{e}_{ph}^{twcc}=2\overline{e}_{ph}(1-\overline{e}_{ph}).
\end{equation}
$n_{t1},n_{t2}$ are the number of survived bits from the bit pairs containing two $0$ bits, two $1$ bits after standard TWCC, and $n_{t3}$ is the number of survived bits from odd-parity bit pairs. $E_1,E_2,E_3$ are the corresponding bit-flip error rates. Those values can be directly observed in the experiment. 

To perform AOPP, Bob actively random pairs the bits $0$ with bits $1$, and Bob get $n_g=min(n_{b0},n_{b1})$ pairs where $n_{b0},n_{b1}$ are the number of bits $0$ and bits $1$ in the raw keys before AOPP. Then Bob announces all the paired sequences to Alice through the public channel. Alice would announce all the positions of the pairs with odd-parities and Alice and Bob only keeps one bits from those announced pairs. The survived bits form a new bit string and would be performed the error correction and privacy amplification to distil the final keys according to the following key rate formulas

\begin{equation}\label{key_rate3}
R^{\prime\prime}=\frac{1}{N}\{n_{u}^{aopp}[1-H(\overline{e}_{ph}^{aopp})]-fn_{t}^{aopp}H(E_{aopp})\}.
\end{equation} 
Here $n_{u}^{aopp}$ is the untagged bits after AOPP and
\begin{equation}
n_{u}^{aopp}=\frac{n_{u0}}{n_{b0}}\frac{n_{u1}}{n_{b1}}n_g.
\end{equation}
$n_{t}^{aopp}$ is the number of survived bits after AOPP and $E_{aopp}$ is the corresponding bit-flip error rate.

\bibliography{refs-jiang}

\end{document}